# Imaging graphene moiré superlattices via scanning Kelvin probe microscopy

Junxi Yu[1,2], Rajiv Giridharagopal[1], Yuhao Li[3], Kaichen Xie[4], Jiangyu Li,[5] Ting Cao[4], Xiaodong Xu,[3] David S. Ginger[1,*]

## ABSTRACT

Moiré superlattices in van der Waals heterostructures are gaining increasing attention because they offer new opportunities to tailor and explore unique electronic phenomena when stacking 2D materials with small twist angles. Here, we reveal local surface potentials associated with stacking domains in twisted double bilayer graphene (TDBG) moiré superlattices. Using a combination of both lateral Piezoresponse Force Microscopy (LPFM) and Scanning Kelvin Probe Microscopy (SKPM), we distinguish between Bernal (ABAB) and rhombohedral (ABCA) stacked graphene and directly correlate these stacking configurations with local surface potential. We find that the surface potential of the ABCA domains is ~15 mV higher (smaller work function) than that of the ABAB domains. First-principles calculations based on density functional theory further show that the different work functions between ABCA and ABAB domains arise from the stacking-dependent electronic structure. We show that, while the moiré superlattice visualized by LPFM can change with time, imaging the surface potential distribution via SKPM appears more stable, enabling the mapping of ABAB and ABCA domains without tip-sample contact-induced effects. Our results provide a new means to visualize and probe local domain stacking in moiré superlattices along with its impact on electronic properties.

[1] Department of Chemistry, University of Washington, Seattle, Washington 98195, United States.
[2] Key Laboratory of Low Dimensional Materials and Application Technology of Ministry of Education, and School of Materials Science and Engineering, Xiangtan University, Xiangtan 411105, Hunan, China.
[3] Department of Physics, University of Washington, Seattle, WA 98195, USA.
[4] Department of Materials Science and Engineering, University of Washington, Seattle, WA 98195, USA.
[5] Department of Materials Science and Engineering, Southern University of Science and Technology, Shenzhen 518055, Guangdong, China. [*] Author to whom the correspondence should be addressed to: dginger@uw.edu.

**Introduction**

Stacking regular 2-dimensional lattices with a precise orientation can give rise to a moiré pattern with a periodicity larger than that of the constituent lattices.[1] Constructing such moiré superlattices using twisted van der Waals heterostructures has recently emerged as an effective approach to engineer a range of fascinating properties such as unconventional superconductivity and Mott-like insulators in magic-angle bilayer graphene superlattices.[2–4] This additional degree of tunability emerges because the electronic properties of 2D materials like graphene are highly sensitive to both the stacking configuration and the layer number.[5–7] For example, a linear dispersion relation between energy and momentum near the K point of the Brillouin zone is observed in monolayer graphene.[8–10] However, in few-layer graphene, interlayer hybridization of the bands is modulated by the moiré superlattice, resulting in modification of the band structure depending on the stacking order.[7,11,12] As an example, the Bernal (ABAB) (Figure 1a) four-layer stacked graphene exhibits two pairs of low-energy bands, whereas the rhombohedral (ABCA) (Figure 1b) stacked four-layer graphene has only one pair of low-energy flat bands.[6,13] These distinctive band structures in turn are predicted to give rise to different transport properties, and ABCA is expected to host a range of correlated and topological states.[14] Identifying the local moiré pattern is thus important for both fundamental and applied studies of stacked van der Waals heterostructures; however, doing so has traditionally been challenging.

Previously, researchers have reported using several microscopy methods to directly image the local moiré pattern, including transmission electron microscopy (TEM)[15,16] and scanning tunnelling microscopy (STM)[17,18]. These methods can atomically resolve the moiré patterns of 2-dimensional heterostructures with the local electronic structure information, albeit at cryogenic temperatures, in ultrahigh vacuum, and require specialized sample preparation. Recently, McGilly et al. employed lateral Piezoresponse Force Microscopy (LPFM)[19] imaging of moiré patterns in stacked bilayer graphene, which probes the piezoelectric response in contact mode. LPFM is convenient because it can be conducted in many commercial atomic force microscopes at room temperature.

In this Letter, we use a combination of both LPFM and scanning Kelvin probe microscopy (SKPM)[20,21] to investigate twisted double-bilayer graphene (TDBG) consisting of two layers of bilayer graphene that are twisted by a small angle. First, we directly establish a correlation between the stacking orders and surface potential, showing alternating triangular ABAB and ABCA domains of different surface potential. Then, using both amplitude modulation (AM) and frequency modulation (FM) SKPM, we show that the surface potential of ABCA is around 15 mV higher than ABAB stacking domains, reflecting their underlying difference in electronic structure. Finally, we observe that LPFM imaging conditions, which require tip-sample contact, can induce slow changes of the moiré superlattices that are likely due to the local mechanical force and oscillating voltage exerted by tip. In contrast, we find that the moiré superlattices are stable under SKPM conditions due to the noncontact mode. In agreement with the SKPM measurements, density-functional theory (DFT) calculations further demonstrate that the work function of the tetralayer graphene system depends upon the interlayer stacking order, which modifies the electronic structure of the system.

**Results and Discussions**

We used the "tear and stack"[22,23] technique to stack two sheets of twisted Bernal stacked bilayer graphene with a small angle.[24–26] This stacking can result in the formation of both Bernal (ABAB) and rhombohedral (ABCA) stacking configurations, as shown in Figures 1a and 1b. At a small twist angle, the lattice energetically prefers to relax into triangular domains with alternating domains of ABAB and ABCA as a result of the atomic reconstruction.[16,27] Figure 1c shows an amplitude image from single-frequency LPFM that is consistent with those reported by McGilly et al.[19] Although single-frequency LPFM can give reliable images on flat surfaces, it can also be sensitive to artifacts,[28] many of which are associated with shifts in the contact resonance frequency induced by the change in tip-sample contact stiffness leading to common misinterpretations of single-frequency measurements.[29] Figure S1a, imaged at the same region of Figure 1c, shows such an example of topography crosstalk.

Accordingly, we also imaged the sample using dual-AC-resonance-tracking (DART) mode LPFM.[30,31] DART mode was developed to reduce crosstalk due to shifts in the contact resonance frequency arising from variations in sample mechanical properties or topography during the PFM measurements.[30,31] While even DART-mode PFM can result in artifacts in some cases due to cantilever beam-induced field effects,[32] it is still generally viewed as more robust as compared to single-frequency PFM modes. Figure 1d shows the resulting DART LPFM amplitude image where the contact resonance frequency has been successfully tracked during scanning with a resonance frequency ranging from 634 kHz to 642 kHz (frequency is shown in Figure 1e). Figure S1b shows

DART LPFM amplitude image measured at the same region of Figures 1d and 1e where DART failed to track the resonance frequency (Figure S1c), and appears similar to Figure S1a. We note that the two PFM amplitude images in Figures 1c and 1d are very similar. This correspondence suggests that the amplitude contrast observed in Figure1c is reliable.

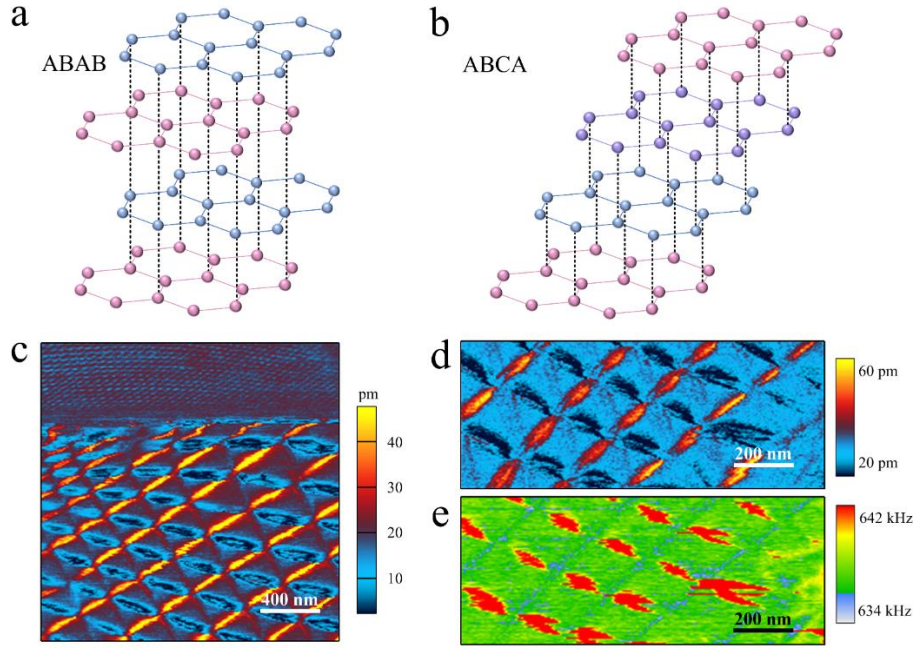

**Figure 1. Single-frequency and DART LPFM imaging of stacking domains in twisted double bilayer graphene.** (a,b) Crystal structure of Bernal ABAB (a) and rhombohedral ABCA (b) stacked twisted double-bilayer graphene. (c) Amplitude response of single-frequency LPFM. (d,e) DART LPFM responses of amplitude (d) and resonance frequency (e).

To characterize the moiré superlattices, we study the TDBG using DART LPFM in ambient conditions. As shown in Figure 2a, there is no moiré superlattice observed in the topography. Figures 2b and 2c show the DART-mode LPFM amplitude and phase images, respectively. Both the amplitude (Figure 2b) and phase images (Figure 2c) show moiré superlattices with separated domains and domain walls, consistent with the expected atomic reconstruction.[16,27] According to previous reports, ABAB domains should have lower energy compared with the ABCA domains, leading to the bending of domain walls into ABCA domains, with the resulting concave ABCA domains.[14,27] In Figures 2b-c the dashed white curves outline the ABAB and ABCA domains where ABAB is convex while ABCA is concave. Additional amplitude and phase images acquired at the second drive frequency of the DART LPFM are shown in Figures S2a and S2b, agreeing well with the data shown in Figures 2b and 2c. Further, these amplitude responses in Figure 2b and Figure S2a are quite close to each other. The resonance frequency (Figure S2c) in the range of 638 kHz to 641 kHz is consistent with the frequency range of Figure 1e, suggesting that DART LPFM results are reliable.

Because of the distinct differences in stacking symmetry between ABAB and ABCA, these two superlattices can exhibit distinct electronic properties. In order to explore possible differences in the electronic properties of the moiré superlattice, we turn to SKPM to measure the local surface potential. Figure 2d shows the resulting surface potential image taken using amplitude modulation mode (AM-SKPM). This image shows that ABCA domains have higher surface potentials than neighboring ABAB domains.

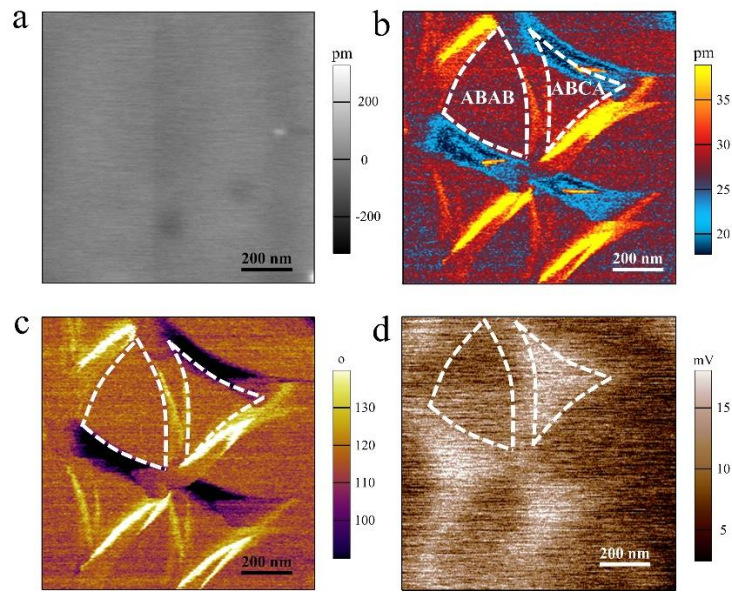

**Figure 2. Correlating stacking configurations with surface potential using a combination of both LPFM and SKPM.** (a) Topography. (b,c) DART LPFM responses of amplitude (b) and phase (c) measured at the same sample area. (d) Corresponding surface potential showing that the concave ABCA regions exhibit ~15 mV higher surface potential (smaller work function) than the convex ABAB domains.

In order to confirm that the surface potential distribution is governed by electronic properties rather than topographic or elastic cross talk, we perform AM-SKPM with varying tip-sample distances on the way through the scan to help rule out topography crosstalk. Before the distance-dependent SKPM measurements, we conducted DART PFM to visualize the moiré superlattice structure (Figure S3a) and the corresponding SKPM images with relative lower tip-sample distance of around 26 nm (Figure S3b), showing the alternating ABAB and ABCA domains of different surface potential, which agrees with Figure 2d. Using these images as a reference, we found that the acquired distance-dependent AM-SKPM, as shown in Figure S3c, also shows alternating triangular patterns of different surface potential consistent with Figure S3b, thereby ruling out the possibility surface topography crosstalk. Thus, we conclude that SKPM can be used to electronically characterize moiré superlattices accurately in a noncontact manner.

The SKPM images in Figure 2 show a difference of ~ 15 mV between ABAB and ABCA domains, with ABCA domains exhibiting the higher surface potential (smaller work function). The SKPM data in Figure 2 was taken with conventional AM-SKPM. AM-SKPM is easily accessible but, because it is sensitive to the electrostatic force between the tip and the sample, it is generally viewed as having lower resolution and accuracy compared to frequency modulation SKPM (FM-SKPM), especially on surfaces with appreciable topography or large/non-periodic features.[33] Thus, we also imaged the surface potential of the moiré superlattices using FM-SKPM. Because FM-SKPM derives its signal from the electrostatic force *gradient* between the tip and sample, it better confines the signal to the tip-sample junction region and should, in theory, possess higher resolution and lead to more accurate measurements.[20] Figure 3a shows the surface potential distribution taken using AM-SKPM while Figure 3b shows the surface potential image taken using FM-SKPM. Figures 3c and 3d show the corresponding surface potential line profiles taken along the same red lines marked in Figures 3a and 3b. Figure 3 shows that, whether imaged with AM-SKPM or FM-SKPM, the very flat TDBG sample shows the same size feature, same contrast, and same signal magnitudes, with the ABCA domains being roughly 15 mV more positive (shallower work function) than the ABAB domains. Recently, Li et al. reported that rhombohedral stacked ABC domain has a higher surface potential than Bernal stacked ABA domain by around 10 mV to 20 mV using SKPM, which Li et al. concluded as indicating that ABA has a higher work function than ABC in trilayer graphene. This magnitude of the potential difference Li et al. measured in trilayer graphene is indeed similar to the magnitude of the potential difference we measure for TDBG. Thus, the question is: what is responsible for the surface potential difference between ABAB and ABCA domains?

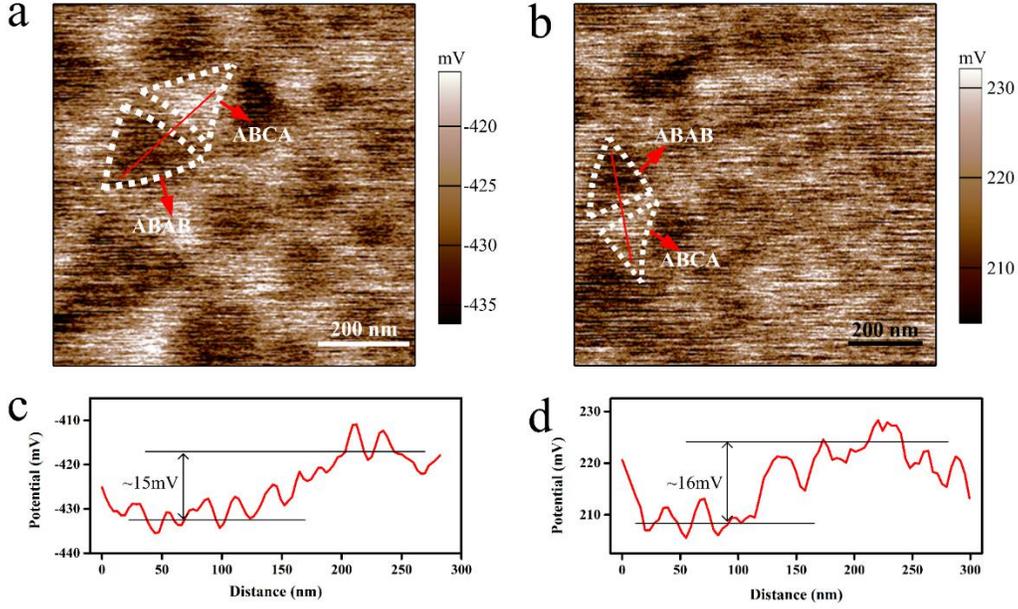

**Figure 3. Surface potential difference between ABAB and ABCA.** (a) AM-SKPM image. (b) FM-SKPM image. (c) Line profile of the surface potential along the red line marked in (a). (d) Line profile of the surface potential along the red line marked in (b).

We note that electronic variations due to differences in stacking orientation might be expected for multiple reasons. We consider two hypotheses: (1) that the change results from different symmetry breaking between the ABAB and ABCA structures, resulting in a change in surface dipole, (2) that the electronic structure differences due to different interlayer hybridization produce an intrinsically lower work function for ABCA graphene. In the first case, a different surface dipole between the ABAB and ABCA domains would alter the measured surface potential, and the experimentally measured shallower work function of the ABCA graphene would suggest a more positive dipole moment pointing out of the surface normal in the ABCA stacking compared to ABAB stacking. In the second case, the differences in stacking may affect the surface potential through its effect on the electronic structure. For example, the electron-hole symmetries of the band structures in the ABAB and ABCA graphene can be broken differently,[34] resulting in a difference in the Fermi energy. The variation in the quantum confinement effects between the ABAB and ABCA stacked graphene can also lead to different electron spatial distribution for states near the Fermi level.

To explore these hypotheses in more detail, we performed DFT calculations on ABAB and ABCA stacked graphene and plot their Kohn-Sham band structures in Figure 4. These two stacking arrangements exhibit distinct electronic band structures where ABCA configurations have a higher Fermi energy than that of the ABAB stacking configurations by ~16 meV, obtained using the local-density approximation (LDA) (~ 21 meV using the generalized gradient approximation, shown in Figure S4). This result is in good agreement with the experimental observations, and also shows consistent trend with previous experiments in trilayer graphene.[27] The main difference between Figure 4a and Figure 4b is that the bands in the vicinity of the Fermi level are very dispersive in the ABAB graphene, whereas the bands in the vicinity of the Fermi level are almost flat in the ABCA graphene because these bands are spatially localized to one sublattice of the top layer and one sublattice of the bottom layers.[35]

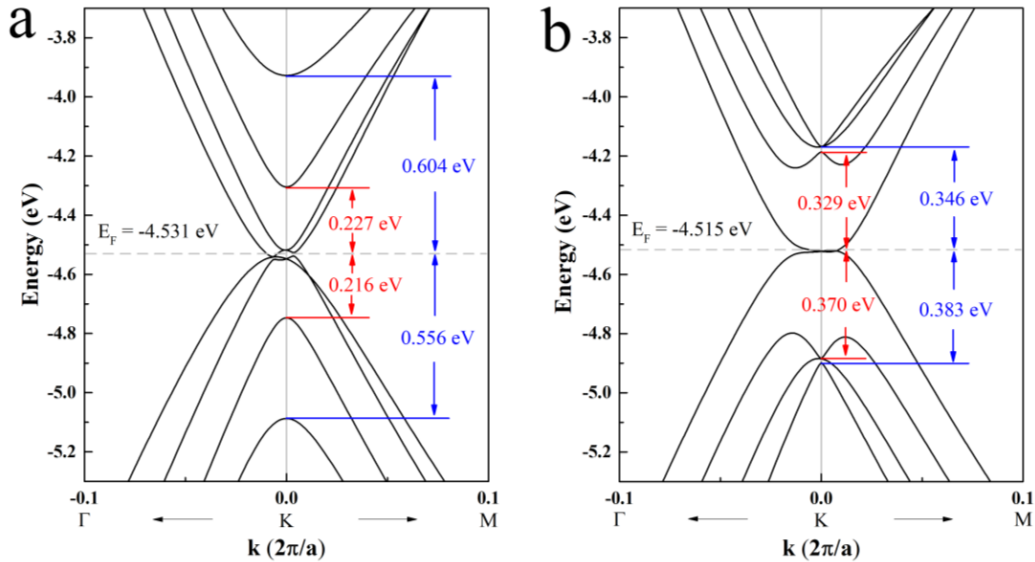

**Figure 4. Calculated DFT-LDA band structures for stacked graphene**. (a) ABAB stacked graphene. (b) ABCA stacked graphene. The Brillouin-zone path is along the Γ-K-M directions from left to right, and it is centered in K. The Kohn-Sham band energies are defined relative to the vacuum level in both cases. From DFT-LDA, the Fermi levels (shown by grey dash line) of ABAB and ABCA stacked graphene are -4.531 eV and -4.515 eV, respectively. A Fermi-Dirac smearing of 25 meV is used for electron occupation.

Returning to the two proposed origins (surface dipoles and electronic structures), our theory and calculations favor hypothesis 2 – an intrinsic and subtle difference in the electronic structure. First, the ABAB and ABCA graphene both have inversion symmetry in their structures. As a result, no intrinsic electric polarization should be expected. (In our experiment, although the different environment at the top and bottom surface could result in a slightly broken inversion symmetry, we do not expect a noticeable shift of Fermi level at room temperature.) For hypothesis 2, we notice that the higher-energy conduction bands and lower-energy valence bands are not located symmetrically relative to the Fermi level, which indicates electron-hole symmetry breaking in both ABAB and ABCA cases. Further scrutiny reveals that in ABCA graphene, the conduction band edge at K is closer to the Fermi level than a corresponding valence band edge at K (labeled by red and blue lines in Figure 4b). This trend is reversed in the ABAB graphene, where the valence bands, rather than conduction bands, are closer to the Fermi level at K (labeled by red and blue lines in Figure 4a). Therefore, the electron-hole symmetry breaking behaves differently in the ABAB and ABCA graphene, which likely causes a change of the Fermi energy. We thus conclude that the intrinsic differences in electronic structures, rather than surface dipoles/symmetry breaking are responsible for the observed small difference in work function between ABAB and ABCA domains.

Finally, we return to the question of image stability. During repeated scanning of the surfaces with PFM, we often noticed that the PFM images would exhibit apparent shifts in the domain structure over periods of tens of minutes (Figure S5). We attribute these shifts to the local mechanical force and sinusoidal voltage exerted by the tip during PFM scanning, or alternatively to the local deposition of water and contamination from the tip-sample contact that inevitably forms when imaging in contact mode in an ambient environment. In contrast, SKPM is a non-contact method, and appears not to perturb the observed domain structure, even after repeated scanning (Figure S6). Given these different time-dependent behaviors, we propose that SKPM will provide advantages for imaging work on moiré superlattices under many conditions.

**Conclusion**

In summary, we use a combination of both LPFM and SKPM to characterize moiré superlattices in twisted double bilayer graphene and show that we can distinguish between the Bernal ABAB and rhombohedral ABCA stacking configurations using either method. Additionally, our results indicate that contact-induced variations in moiré superlattices may occur in LPFM and therefore convolve mechanical force effects with stacking order. By correlating DART-mode LPFM with SKPM, we find that the surface potential distribution is directly correlated with the stacking order. The surface potential of ABCA domains is higher (the work function is smaller) than that of the ABAB domains by ~15 mV. DFT calculations including electron correlation at the mean-field level are consistent with the experimental result, indicating that ABCA has a smaller work function. Our results provide a new route to visualize stacking-induced surface potential distributions, as well as provide a direct probe of the electronic consequences of local domain stacking.

## Methods

**Lateral piezoelectric force microscopy and scanning kelvin probe microscopy.** An Asylum Research MPF-3D atomic force microscopy was used for LPFM and SKPM studies in ambient condition. FM-KPFM code is available upon request.[36] The ElectriMulti75-G conductive probe was made of silicon with the tip and cantilever coated with platinum, and the radius of the tip was less than 30 nm. The free resonance frequency of the probe is ~70 kHz and the spring constant of the probe was ~ 2.8 N/m.

**Sample fabrication.** The fabrication of TDBG sample followed by a standard 'tear and stack' technique. A polycarbonate (PC) film on tip of polydimethyl siloxane stamp was used to sequentially pick up hexagonal boron nitride, then half of bilayer graphene, followed by the second half of bilayer graphene with a small twist angle. This structure was flipped over and placed on a $Si/SiO_2$ wafer.

**First-Principles Calculations.** Ab initio calculations were performed using DFT within the LDA exchange-correlation functional, implemented in the Quantum Espresso package.[37] Norm-conserving pseudopotentials were employed with a plane-wave energy cutoff of 80 Ry. The structures were fully relaxed until the force on each atom was <0.005 eV/Å. The calculated lattice constant is a = 2.45 Å for both ABAB and ABCA stacking configurations. The calculated interlayer distance is d = 3.33 Å for both ABAB and ABCA stacked graphene. A 100×100 k-grid is used for Brillouin zone sampling. A vacuum region of 15.0 Å was used in the out-of-plane direction to avoid interaction between periodic images. Results using the GGA function is included in the supplementary materials.

## ACKNOWLEDGEMENTS

This paper is based primarily on work supported by the NSF MRSEC program under a SuperSeed award (1719797). J.Y. thanks the support of the China Scholarship Council for a fellowship. K.X.

acknowledges the Graduate Fellowship from Clean Energy Institute funded by the State of Washington.**REFERENCES**

(1) Geim, A. K.; Grigorieva, I. V. Van Der Waals Heterostructures. *Nature* **2013**, *499*, 419–425.

(2) Cao, Y.; Fatemi, V.; Fang, S.; Watanabe, K.; Taniguchi, T.; Kaxiras, E.; Jarillo-Herrero, P. Unconventional Superconductivity in Magic-Angle Graphene Superlattices. *Nature* **2018**, *556*, 43–50.

(3) Cao, Y.; Fatemi, V.; Demir, A.; Fang, S.; Tomarken, S. L.; Luo, J. Y.; Sanchez-Yamagishi, J. D.; Watanabe, K.; Taniguchi, T.; Kaxiras, E.; Ashoori, R. C.; Jarillo-Herrero, P. Correlated Insulator Behaviour at Half-Filling in Magic-Angle Graphene Superlattices. *Nature* **2018**, *556*, 80–84.

(4) He, M.; Li, Y.; Cai, J.; Liu, Y.; Watanabe, K.; Taniguchi, T.; Xu, X.; Yankowitz, M. Tunable Correlation-Driven Symmetry Breaking in Twisted Double Bilayer Graphene. *Nat. Phys.* **2020**, *17*, 26–30.

(5) Yankowitz, M.; Wang, J. I. J.; Birdwell, A. G.; Chen, Y. A.; Watanabe, K.; Taniguchi, T.; Jacquod, P.; San-Jose, P.; Jarillo-Herrero, P.; LeRoy, B. J. Electric Field Control of Soliton Motion and Stacking in Trilayer Graphene. *Nat. Mater.* **2014**, *13*, 786–789.

(6) Latil, S.; Henrard, L. Charge Carriers in Few-Layer Graphene Films. *Phys. Rev. Lett.* **2006**, *97*, 036803.

(7) Castro Neto, A. H.; Guinea, F.; Peres, N. M. R.; Novoselov, K. S.; Geim, A. K. The Electronic Properties of Graphene. *Rev. Mod. Phys.* **2009**, *81*, 109–162.

(8) K. S. Novoselov, A. K. Geim, S. V. Morozov, D. Jiang, Y. Zhang, S. V. Dubonos, I. V. G. and A. A. F. Electric Field Effect in Atomically Thin Carbon Films. *Science* **2004**, *306*, 666–669.11